\documentclass{article}
\usepackage{spconf,amsmath,graphicx}
\usepackage[numbers]{natbib}

\usepackage{enumitem}
\setlist{nosep, leftmargin=14pt}
\usepackage{amsmath,amssymb,amsfonts}

\usepackage{mwe} % to get dummy images

\title{Cortical Surface Diffusion Generative Models}
\name{Zhenshan Xie, Simon Dahan, Logan Z. J. Williams, M. Jorge Cardoso, Emma C. Robinson}
\address{King's College London\\
School of Biomedical Engineering $\And$ Imaging Science}
\begin{document}
%\ninept
%
\maketitle
\begin{abstract}
Cortical surface analysis has gained increased prominence, given its potential implications for neurological and developmental disorders. Traditional vision diffusion models, while effective in generating natural images, present limitations in capturing intricate development patterns in neuroimaging due to limited datasets. This is particularly true for generating cortical surfaces where individual variability in cortical morphology is high, leading to an urgent need for better methods to model brain development and diverse variability inherent across different individuals. In this work, we proposed a novel diffusion model for the generation of cortical surface metrics, using modified surface vision transformers as the principal architecture. We validate our method in the developing Human Connectome Project (dHCP) with results suggesting that our model demonstrates excellent performance in capturing the intricate details of evolving cortical surfaces. Furthermore, our model can generate high-quality realistic samples of cortical surfaces conditioned on postmenstrual age (PMA) at scan.

\end{abstract}

%% Proof of principle
%% Future work aims for translation of data based on different conditions
%% Generative model
%%% Novel contributions
%%%% Validating this on real-world medical dataset 
%%%% Incorporated transformer architecture
%%%% First diffusion model on cortical surface -> diffusion in general for cortical surfaces 

%% Uncondition - generate a brain that looks realistic
%% Condition on age 
%% Can an SiT be fooled by looking at synthetic vs. real data 

\section{Introduction}
\label{sec:intro}
The human cerebral cortex is a highly-convoluted sheet of grey matter that is best modeled as a surface ~\cite{fischl1999high,robinson2014msm,glasser2016human}. As one of the most highly evolved areas of the brain relative to non-human primates, cortical function underpins many aspects of higher-order cognition and is implicated in many neurological and psychiatric disorders \cite{paus2008many,roe2021asymmetric}. Unfortunately, localising signs of pathology in individual brains can be extremely challenging due to the scale of variation of human cortical organisation ~\cite{glasser2016multi,kong2019spatial,gordon2017precision}, which obscures subtle disease. Therefore, a primary goal of translational neuroimaging is to develop methods that accurately disentangle healthy variation from pathology. 

%%% Cerebral cortex varies a lot, need good methods. 
One approach that has gained recent prominence is normative modeling. Here the objective is to learn a generative model that encodes how brains vary normally relative to continuous phenotypes such as age or behavioural scores, from which signs of disease may be detected as outliers from the model \cite{marquand2016understanding,rutherford2022normative}. Classically normative models of cortical variation have been fit with Gaussian Process Regression (GPR) \cite{marquand2016understanding,rutherford2022normative} or Generative Additive Models of Location Scale and Shape (GAMLSS) ~\cite{bethlehem2022brain}. However, these require hand engineering of imaging features, generally using traditional image processing pipelines that combine diffeomorphic image registration with label propagation of cortical areas from population average templates. Approaches that have been shown repeatedly to under-estimate the full scale of cortical variation ~\cite{glasser2016human}.

A potentially more powerful approach is to use deep generative models to encode normal brain variation and disentangle it from disease ~\cite{bass2022icam}. To this end, denoising Diffusion Probabilistic Models (DDPM)~\cite{ho2020denoising} have recently emerged as extremely powerful generators of natural images. Several works~\cite{liu2023meshdiffusion,huang2022riemannian} have already gained great success in adapting diffusion models to non-Euclidean data, using Stochastic Differential Equations (SDE) to approximate samples on manifolds, but focusing on simple mesh and manifolds generation. However, modeling cortical surfaces requires using high-resolution grids which are not feasible with current methodologies. 

%no models typically struggle to model complex distributions, %like cortical surfaces, 
%facing problems like model collapse% and the challenge of learning expressive features by SDEs. 
To address this problem, we chose to build from recent work on the development of surface vision transformers (SiT)~\cite{dahan2022}. These have been validated as powerful encoders of cortical feature maps, outperforming surface convolutional approaches to demonstrate state-of-the-art performance of cortical phenotype regression tasks ~\cite{dahan2022}. In this paper we demonstrate that, with the use of an SiT as a backbone for DDPM generation, it is possible to conditionally generate feature maps that accurately simulate cortical neurodevelopment. 
\begin{figure*}[!t]
    \centering
    \includegraphics[width=\textwidth,height=0.3\textwidth]{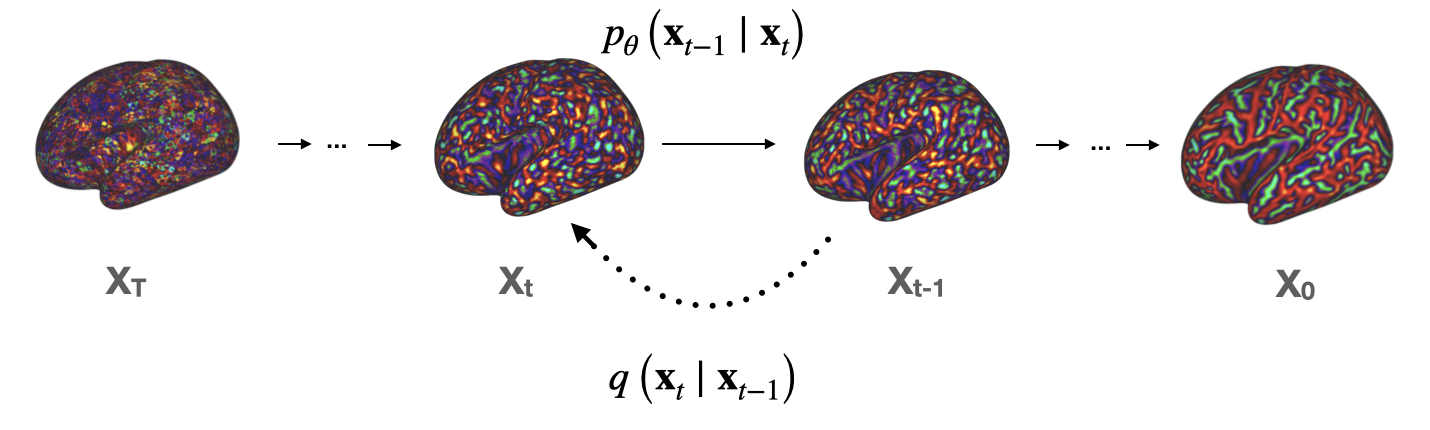}
    \caption{The directed graphical model of diffusion process on cortical surface}
    \label{fig:your-label}
\end{figure*}

\section{Method}
\label{sec:format}

\subsection{DDPM formulation}
The basis of a DDPM ~\cite{ho2020denoising} is a sequential diffusion model. This is essentially a forward diffusion process that operates as a Markov chain, incrementally applying Gaussian noise with variance $\beta_t \in(0,1)$ to the original data distribution $q\left(x_0\right)$. The forward transformation can be represented as follows: 
\begin{equation}
q\left(x_{1: T} \mid x_0\right)  =\prod_{t=1}^T q\left(x_t \mid x_{t-1}\right)
\end{equation}
\begin{equation}
\label{eq:add_noise}
    q\left(x_t \mid x_{t-1}\right)  =\mathcal{N}\left(x_t ; \sqrt{1-\beta_t} x_{t-1}, \beta_t \boldsymbol{I}\right)
\end{equation}
With $x_0$ representing the initial data, and \\$x_{1: T}=\{x_1, x_2, \ldots, x_T\}$ representing the latent variables, generated from incremental addition of Gaussian noise (Eq \ref{eq:add_noise}). Generation of the diffusion model is learning to reverse the process by using a neural network ($\epsilon_\theta$) to iteratively denoise samples over time-steps $t$. The output of the model is a prediction of the statistics of the distribution: $p_\theta\left(x_{t-1} \mid x_t\right)$:
\begin{equation}
    p_\theta\left(x_{t-1} \mid x_t\right)=\mathcal{N}\left(\mu_\theta\left(x_t\right), \Sigma_\theta\left(x_t\right)\right)
\end{equation}
This model is trained with the variation lower bound of the log-likelihood of $x_0$, reducing:
\begin{equation}
\begin{aligned}
     \mathcal{L}(\theta)=&-p\left(x_0 \mid x_1\right)+\\
     \sum_t \mathcal{D}_{K L}&\left(q\left(x_{t-1} \mid x_t, x_0\right)|| p_\theta\left(x_{t-1} \mid x_t\right)\right)
\end{aligned}
\end{equation}
As both $q$ and $p_\theta$ are Gaussian noise, $ \mathcal{D}_{K L}$ can be evaluated from the mean and covariance of two distributions. So the noise prediction network $\epsilon_\theta$ can be trained easily from the mean-squared error between the predicted noise $\epsilon_\theta\left(x_t\right)$ and the sampled Gaussian noise $\epsilon_t$. To generate a sample through the process, an initial point is drawn from $p\left(x_T\right)$ with the same dimensions as the original training data $x_0$. Then samples $\{x_{T-1},x_{T-2}...{x_1}\}$ are sequentially denoised by the $\epsilon_\theta$ until the model approximates the original data point $x_0$. %In our work, the sample will be replaced by surface and details will be shown in the following sections. 

\subsection{Surface Diffusion Model}
In this paper, samples $x_0\in \mathbb{R}^{40962\times 1}$ represent cortical curvature maps, projected to a regular, sixth-order icospheric grid 40962 regularly spaced vertices. % $I_6=(V_6,F_6)$, with $|V_6|=40962$ vertices and $|F_6|=81920$ faces. The reverse diffusion process is parameterised using a surface vision transformer (SiT) \cite{dahan2022}.
% which translated the concept of ViTs to sphericalised cortical meshes by representing the cortical surface as a mesh structure on a sphere, to be the neural network for noise predictor in the diffusion process.
% In this section, we describe the components of the modified SiT.
\begin{figure}[h]
    \centering
    \includegraphics[width=0.48\textwidth, height=0.35\textwidth]{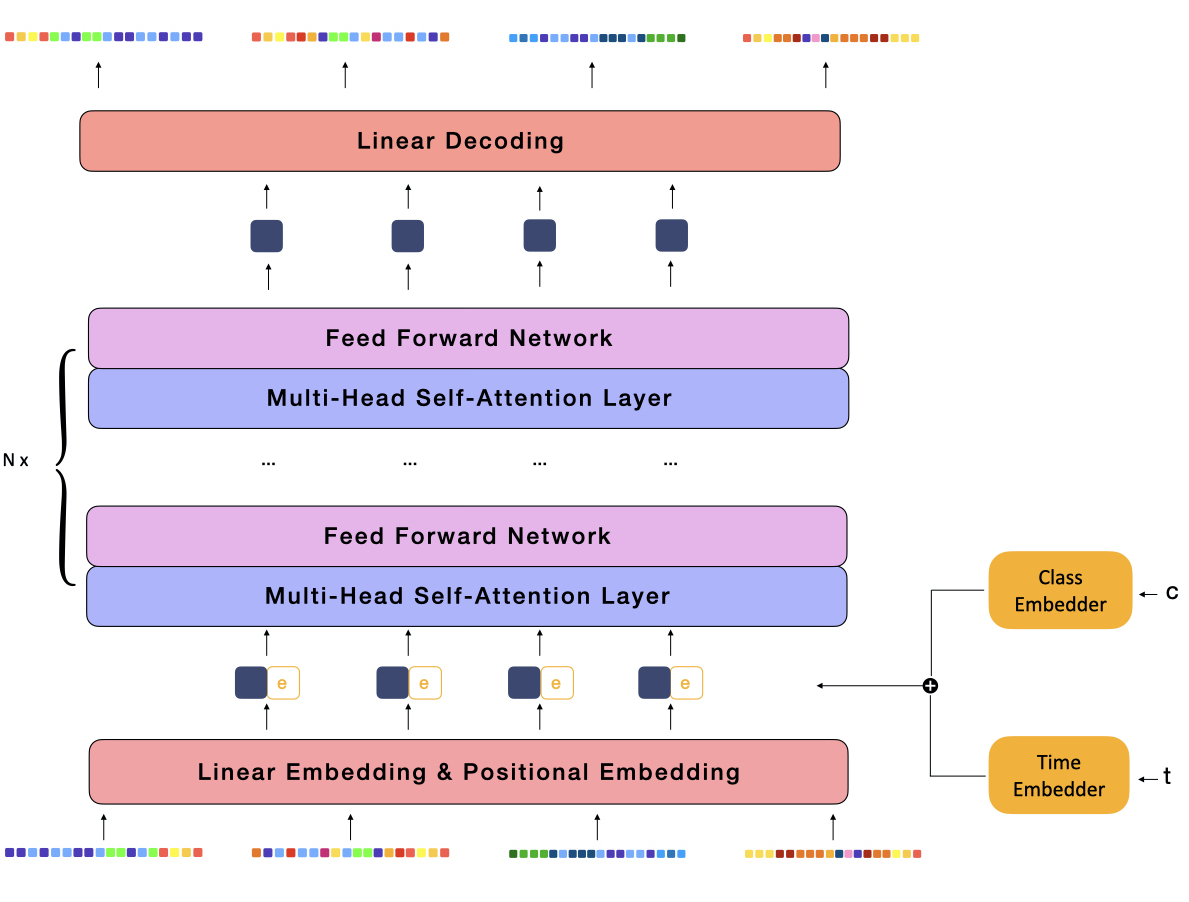}
    \caption{The figure shows the architecture of the cortical surface diffusion model. First, the flattened input will be projected by linear embedding. Then, positional embedding and condition embeddings (class embedding and time embedding) would be added to the input tokens. After N transformer blocks, the prediction of noise would be projected out by the linear decoding layer. }
    \label{block}
\end{figure}
%\subsection{Surface Vision Transformer (SiT)}
\subsubsection{Surface Vision Transformer} The SiT \cite{dahan2022} translates surface understanding to a sequence-to-sequence learning paradigm by tessellating cortical feature maps (sampled on a regular high resolution spherical icosahedral grid) according to regular triangular subdivisions of a lower resolution icosahedron. In this paper, we use ico6 to represent the data, and ico2 ($I_2=(V_2,F_2)$, $|V_2|=162$ vertices and $|F_2|=320$) for sampling. Each patch is generated from the intersection of vertices ($V_p$) from ico6 that overlap with each face of ico2, which results in a sequence of 320 patches, each of $V_p =153$ vertices. This generates a sequence of $|F_2|$ flattened patches of length $V_p$: $\widetilde{X}=\left[\widetilde{X}_1^{(0)},\ldots,\widetilde{X}_{|F_2|}^{(0)}\right]\in R^{|F_2| \times V_p}$, which is first projected onto a D-dimensional sequence of tokens $X^{(0)}=\left[X_1^{(0)}, \ldots, X_{|F_2|}^{(0)}\right] \in R^{|F_2| \times D}$, using a trainable linear layer (see Figure \ref{block}). Positional embedding is then added to encode spatial information about the sequence of tokens. The data is then passed through $N$ vision transformer blocks, each composed of a multi-head self-attention layer and a feed-forward network (for more details of implementation see \cite{dahan2022}).

\subsubsection{Decoding}
In its original form, the SiT was designed for classification or regression \cite{dahan2022}. To support iterative prediction of noise and diagonal covariance, at the same resolution as the original data, we extend the network with %a linear decoding layer after the final transformer block. %to predict noise and diagonal covariance. 
% As in \cite{dahan2023surface}, the decoder employs an asymmetric design. 
%Here, the linear decoding module applies 
a layer norm (adaptive if using adaLN) and a standard linear layer to decode each token into $N \times(2V_p)$. Finally, the decoded tokens are rearranged into their original spatial layout to regenerate surface maps.

\subsubsection{Conditioning}
Following the DDPM formulation in the previous section, samples are iteratively denoised through the SiT, with noise at each iteration parametrized by iteration parameter $t$. At the same time, we seek to further condition the model on the postmentrual
age of our samples $a$. This is achieved by appending $a$ and $t$ as additional tokens in the input sequence: adding them into both surface input tokens before the SiT (see Figure \ref{block}). 

%\subsection{Generation}
Conditional generation of new samples is supported through use of classifier-free guidance~\cite{ho2022classifier}. This approach requires both conditioned $p_\theta(\mathbf{x} \mid c)$ and unconditioned  $p_\theta(\mathbf{x})$ predictions to be generated from the same network (i.e. parametrized as $\boldsymbol{\epsilon}_\theta\left(\mathbf{x}_t, t, c\right)$ and $\boldsymbol{\epsilon}_\theta\left(\mathbf{x}_t, t\right)$)  allowing
%incorporates scores from a conditional and an unconditional diffusion model, which are learnt via a single network, parameterised with  
%Firstly, we let the unconditional diffusion model $p_\theta(\mathbf{x})$, parameterized with score estimator: $\boldsymbol{\epsilon}_\theta\left(\mathbf{x}_t, t\right)$, and the conditional model $p_\theta(\mathbf{x} \mid c)$ parameterized through $\boldsymbol{\epsilon}_\theta\left(\mathbf{x}_t, t, c\right)$. So 
the gradient of an implicit classifier to be represented as follows:
\begin{equation}
\nabla_{\mathbf{x}_t} \log p\left(c \mid \mathbf{x}_t\right)=\nabla_{\mathbf{x}_t} \log p\left(\mathbf{x}_t \mid c\right)-\nabla_{\mathbf{x}_t} \log p\left(\mathbf{x}_t\right)
\end{equation}
In this setting, classifier-free guidance can be used to encourage the sampling procedure to find $x$ such that $\log p(c \mid x)$ is high. This method enables the surface diffusion model to generate target samples without module modification and the results have shown its performance on surface data.

\section{Experiments and Results}
\label{sec:pagestyle}

%The paper title (on the first page) should begin 1.38 inches (35 mm) from the top edge of the page, centered, completely capitalized, and in Times 14-point, boldface type.  The authors' name(s) and affiliation(s) appear below the title in capital and lower case letters.  Papers with multiple authors and affiliations may require two or more lines for this information.

\subsection{Data}
Data from the developing Human Connectome Project (dHCP) consists of cortical surface meshes and metrics (sulcal depth, curvature, cortical thickness, and T1w/T2w myelination) derived from T1- and T2-weighted magnetic resonance images (MRI) using the dHCP structural pipeline~\cite{Makropoulos2018}. We use 530 scans from term (born after $37$ weeks gestation) and preterm neonates (born before $37$ weeks), covering postmenstrual age (PMA) at scan between 24 to 45 weeks.
%and 111 preterm neonates (born prior to $37$ weeks gestation). 
%95 preterm neonates were scanned twice, once shortly after birth, and once at term-equivalent age.
In this experiment, we only use the curvature map. %The whole training dataset contains 411 scans with their gestational age(GA). 
Training, test, and validation sets were allocated in the ratio of 423:54:53 examples.

\begin{figure}[h]
    \centering
    \includegraphics[width=0.48\textwidth, height=0.3\textwidth]{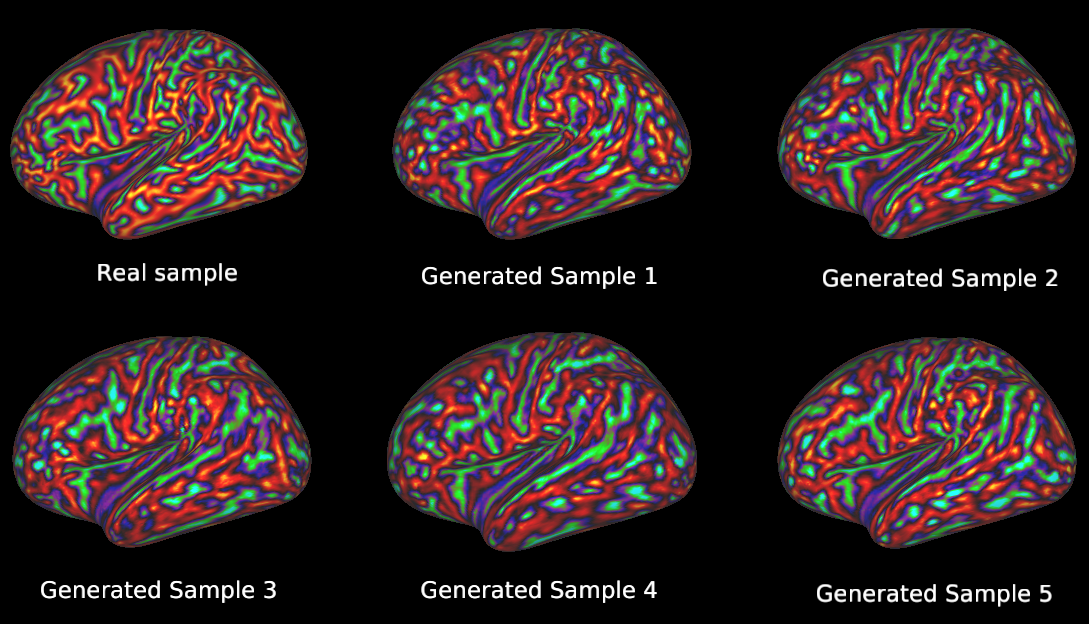}
    \caption{Unconditional generated curvature surface compared with real sample}
    \label{unconditional}
\end{figure}

\subsection{Implementation Details}
\textbf{Training} The SiT transformer backbone was implemented with the same size of SiT-small from ~\cite{dahan2022}, 12 layers, 6 heads, 384 hidden sizes, and 768 Multi-Layer Perception(MLP) sizes. We train both unconditional and conditional diffusion models with AdamW~\cite{loshchilov2017decoupled,kingma2014adam} using standard weight initialization techniques from SiT, except for the final linear (decoding) layer, which is initialised with zeros. Training was implemented with a constant learning rate of $1\times10^{-4}$, no weight decay, and a batch size of 180 on NVIDIA 3090 with 24 GB memory. Following common practice in the generative modeling literature, we maintain an exponential moving average (EMA) of surface diffusion weights with a training decay of 0.9999.

\subsection{Evaluation of Conditional Generation}
\begin{figure}
    \centering
    \includegraphics[width=0.49\textwidth, height=0.525\textwidth]{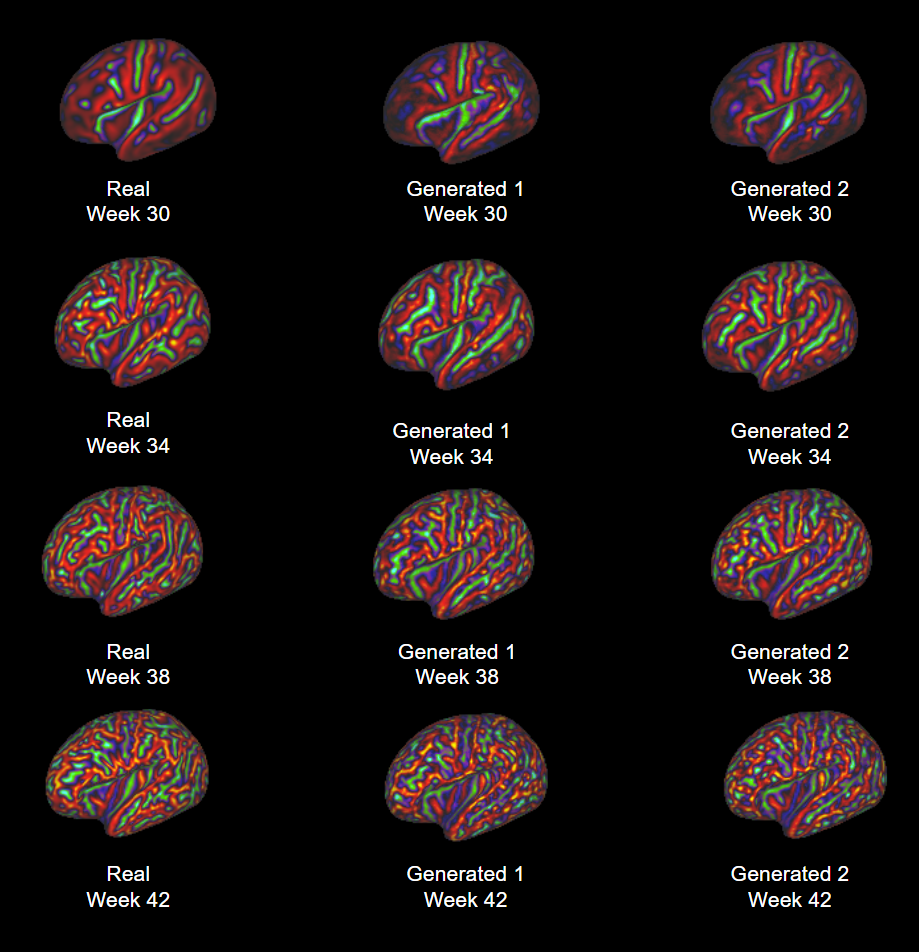}
    \caption{Samples of cortical surfaces generated by surface diffusion model conditioned on birth age of 30, 34, 38 and 42 weeks, compared to the real samples in dHCP dataset(First line)}
    \label{conditional}
\end{figure}

The model was validated through visual inspection and training of an independent SiT regression model which evaluated whether the generated samples convincingly represent curvature maps of a specific PMA. This regression model was trained on the same data set used for generation, with PMA as the target variable. The proposed surface DDPM was used to generate 20 synthetic examples for each week, from 27 weeks to 44 weeks. The results in table \ref{table:regresssionerror} report mean absolute error for PMA predictions obtained on ground truth testing data and on synthetic data. Visual examples are shown in Fig \ref{conditional}. Together these results suggest that the proposed model is generating highly realistic cortical curvature maps that can be realistically conditioned on PMA. %Besides, the generated surfaces (shown in Fig.3 and Fig.4) can accurately model the cortical maturation patterns in both preterm and term neonates.
\begin{table}[h]
    \centering
    \begin{tabular}{lcc}
        \hline & Ground Truth & Synthetic \\
        \hline MAE & $0.59 \pm 0.51 $ & $ 0.80 \pm 0.50 $  \\
        r2 & $ 0.96$ & $0.96 $ \\
\hline
\end{tabular}
\label{table:regresssionerror}
\caption{Prediction results on dHCP test and synthetic data for the PMA regression task. Mean Absolute Error (MAE) with standard deviation are reported, as well as $R^2$ score.}
\end{table}

\section{Conclusion}
\label{sec:typestyle}
This paper presents a general cortical surface diffusion generative model validated on conditioned generation of cortical curvature maps across neurodevelopment. The generated samples demonstrate that surface diffusion can capture expressive features on the surface sufficiently well to sensitively model the rapid changes in cortical curvature that occur across late gestation. One major contributing factor is the incorporation of a surface vision transformer model, which has previously been shown to outperform surface convolution for cortical phenotype regression. While the results demonstrate considerable potential, the model in its current form is limited to synthesis of entirely novel samples and cannot intervene, or translate the class of individual examples. Future work will look at the towards frameworks for image-to-image translation \cite{bass2022icam,fawaz2022deep,fawaz2023surface} and deep causal modelling \cite{sanchez2022diffusion}. This would create opportunities for the development of networks suited to anomaly detection of focal cortical malformations (such as epileptogenic lesions \cite{spitzer2023robust}), modelling of cortical atrophy in dementia \cite{bass2022icam}, or investigation of the subtle impacts of preterm birth on cortical development \cite{fawaz2022deep,fawaz2023surface}. 

% In addition, due to the flexibility and scalability of surface transformer, the surface diffusion model can be extended to any other modality surface data. Future work could be focused on the extension of surface diffusion model to different application tasks (neurodevelopmental translation, segmentation, anomaly detection, etc).

%\vfill
%\pagebreak

\section{Acknowledgments}
\label{sec:acknowledgments}
We would like to acknowledge funding from King's-China Scholarship(K-CSC) PhD Scholarship program. Data were provided by the developing Human Connectome Project. We are grateful to the families who generously supported this trial. 
\bibliographystyle{IEEEbib}
\bibliography{strings,refs}

\end{document}